\documentclass{article}

\usepackage{arxiv}

\usepackage[utf8]{inputenc} 
\usepackage[T1]{fontenc}    
\usepackage{hyperref}       
\usepackage{url}            
\usepackage{booktabs}       
\usepackage{amsfonts}       
\usepackage{nicefrac}       
\usepackage{microtype}      
\usepackage{lipsum}
\usepackage{graphicx}
\graphicspath{ {./images/} }

\title{Personalized Early Stage Alzheimer’s Disease Detection: A Case Study of President Reagan’s Speeches\thanks{This paper has been accepted for publication in ACL workshop on BioNLP 2020.}}

\author{
 Ning Wang \\
  Department of Electrical and Computer Engineering\\
  Stevens Institute of Technology\\
  Hoboken, NJ 07030 \\
  \texttt{nwang7@stevens.edu} \\
  \And
 Fan Luo \\
  Department of Computer Science\\
  Stevens Institute of Technology\\
  Hoboken, NJ 07030 \\
  \texttt{fluo40@stevens.edu} \\
  \And
 Vishal Peddagangireddy\thanks{This work was done when Vishal Pendangangireddy was with the Stevens Institute of Technology.}\\
  Department of Electrical and Computer Engineering\\
  Stevens Institute of Technology\\
  Hoboken, NJ 07030 \\
  \texttt{vpeddaga@stevens.edu} \\
  \And
  K.P. Subbalakshmi \\
  Department of Electrical and Computer Engineering\\
  Stevens Institute of Technology\\
  Hoboken, NJ 07030 \\
  \texttt{ksubbala@stevens.edu@} \\
  \And
  R. Chandramouli \\
  Department of Electrical and Computer Engineering\\
  Stevens Institute of Technology\\
  Hoboken, NJ 07030 \\
  \texttt{mouli@stevens.edu} \\
}

\begin{document}
\maketitle

\begin{abstract}
Alzheimer's disease (AD)-related global healthcare cost is estimated to be  \$1 trillion by 2050. Currently, there is no cure for this disease; however, clinical studies show that early diagnosis and intervention helps to extend the quality of life and inform technologies for personalized mental healthcare. Clinical research indicates that the onset and progression of Alzheimer's disease lead to dementia and other mental health issues. As a result, the language capabilities of patient start to decline. 

In this paper, we show that machine learning-based unsupervised clustering of and anomaly detection with linguistic biomarkers are promising approaches for  intuitive visualization and personalized early stage detection of Alzheimer's disease. We demonstrate this approach on 10 year's (1980 to 1989) of President Ronald Reagan's speech data set. Key linguistic biomarkers that indicate early-stage AD are identified. Experimental results show that Reagan had early onset of Alzheimer's sometime between 1983 and 1987. This finding is corroborated by prior work that analyzed his interviews using a statistical technique. The proposed technique also identifies the exact speeches that reflect linguistic biomarkers for early stage AD.

\end{abstract}

\section{Introduction}

Alzheimer’s disease is a serious mental health issue faced by the global population. About 44 million people worldwide are diagnosed with AD. The U.S. alone has 5.5 million AD patients. According to the Alzheimer’s association the total cost of care for AD is estimated to be \$1 trillion by 2050. There is no cure for AD yet; however, studies have shown that early diagnosis and intervention can delay the onset. 

Regular mental health assessment is a key challenge faced by the medical community. This is due to a variety of reasons including social, economic, and cultural factors. Therefore, Internet based technologies that unobtrusively and continually collect, store, and analyze mental health data are critical. For example, a home smart speaker device can record a subject’s speech periodically, automatically extract AD related speech or linguistic features, and present easy to understand machine learning based analysis and visualization. Such a technology will be highly valuable for personalized medicine and early intervention. This may also encourage people to sign-up for such a low cost and home-based AD diagnostic technology. Data and results of such a technology will also instantly provide invaluable information to mental health professionals. 

Several studies show that subtle linguistic changes are observed even at the early stages of AD. In \cite{ForEtal05}, more than 70\% of AD patients scored low in a picture description task. Therefore, a critical research question is:  can spontaneous temporal language impairments caused by AD be detected at an early stage of the disease? Relation between AD, language functions and language domain are summarized in \cite{SzaEtal15}.  In \cite{VenEtal08}, a significant correlation between the lexical attributes characterising residual linguistic production and the integrity of regions of the medial temporal lobes in early AD patients is observed. Therefore, in this paper, we explore a machine-learning based clustering and data visualization technique to identify linguistic changes in a subject over a time period. The proposed methodology is also highly personalized since it observes and analyzes the linguistic patterns of each individual separately using only his/her own linguistic biomarkers over a period of time. 

First, we explore a machine learning algorithm called t-distributed stochastic neighbor embedding (t-SNE) \cite{MaaEtal08}. t-SNE is useful in dimensionality reduction suitable for visualization of high-dimensional datasets. It calculates the probability that two points in a high-dimensional space are similar, computes the corresponding probability in a low-dimensional space, and minimizes the difference between these two probabilities for mapping or visualization. During this process, the sum of Kullback-Leibler divergences \cite{LiuEtal03} over all the data points is minimized. Our hypothesis is that high-dimensional AD-related linguistic features when visualized in a low-dimensional space may quickly and intuitively reveal useful information for early diagnosis. Such a visualization will also help individuals and general medical practitioners (who are first points of contact) to assess the situation for further tests.
Second, we investigate two unsupervised machine learning techniques, one class support vector machine (SVM) and isolation forest, for temporal detection of linguistic abnormalities indicative of early-stage AD. These proposed approaches are tested on President Reagan's speech dataset and corroborated with results from other research in the literature.

This paper is organized as follows. Background research is discussed in Section~\ref{sec:Background}, Section~\ref{sec:dataProcessAndFeatureSelect} presents the Reagan speech data set used in this paper, data pre-processing techniques that were applied and AD-related linguistic feature selection rationale and methodology, Section~\ref{sect:clusterAndVisual} contains the clustering and visualization of the hand-crafted linguistic features and anomaly detection to infer the onset of AD and to detect the time period of changes in the linguistic statistical characteristics, and experimental results to demonstrate the proposed method. In Section~\ref{sect:adad}, we describe the machine learning algorithms for detecting anomalies from personalized linguistic biomarkers collected over a period of time to identify early-stage AD. Concluding remarks are given in Section~\ref{sect:conc}.

\section{Background}
\label{sec:Background}
The “picture description” task has been widely studied to differentiate between AD and non-AD or control subjects. In this task, a picture (the “cookie theft picture”) is shown and the subject is asked to describe it. It has been observed that subjects with AD usually convey sparse information about the picture ignoring expected facts and inferences \cite{GilEtal96}. AD patients have difficulty in naming things and replace target words with simpler semantically neighboring words. Understanding metaphors and sarcasm also deteriorate in people diagnosed with AD \cite{RapEtal11}.  

Machine learning based classifier design to differentiate between AD and non-AD subjects is an active area of research. A significant correlation between dementia severity and linguistic measures such as confrontation naming, articulation, word-finding ability, and semantic fluency exists. Some studies have reported a 84.8 percent accuracy in distinguishing AD patients from healthy controls using temporal and acoustic features (\cite{RenEtal14}; \cite{FraEtal16}).   

Our study differs from the prior work in several ways. Prior work attempt to identify linguistic features and machine learning classifiers for differentiating between AD and non-AD subjects from a corpus (e.g., Dementia Bank \cite{Mac07}) containing text transcripts of interviews with patients and healthy control. In this paper, we first analyze the speech transcripts (over several years) of a single person (President Ronald Reagan) and visualize time-dependent linguistic information using t-SNE. The goal is to identify visual clues for linguistic changes that may indicate the onset of AD. Note that such an easy to understand visual representation depicting differences in linguistic patterns will be useful to both a common person and a general practitioner (who is the first point of contact for majority of patients). Since most general practitioners are not trained mental health professionals the impact of such a visualization tool will be high, especially at the early stages of AD. Sigificant AD-related linguistic biomarkers derived from t-SNE analysis are then used in two unsupervised clustering algorithms for detecting AD-related temporal linguistic anomalies. This provides an estimate for the time when early-stage AD symptoms are beginning to be observed.

\section{Reagan Speeches: Data Collection, Pre-processing, and Feature Selection}
\label{sec:dataProcessAndFeatureSelect}
We describe the data collection, pre-processing, and feature engineering methodologies in this section. Ronald Reagan was the 40th president (served from 1981 to 1989) of the United States of America. He was an extraordinary orator, a radio announcer, actor, and the host for a show called “General Electric Theatre.” Clearly, for being successful in these professional domains one needs to have good memory, consciousness, intuition, command over the language, and ease of communicating with a large audience. Reagan officially announced that he had been diagnosed with AD on November 5, 1994. But it was speculated that his cognitive abilities where on the decline even while in office \cite{GotEtal88}. Therefore, analyzing his speech transcripts for early signs of AD may reveal interesting patterns, if any. 

The Reagan Library is the repository of presidential records for President Reagan’s administration. We download his 98 speeches from 1980 to 1989 as shown in Table~\ref{tab:speeches}. We removed special characters, tags, and numbers and kept only the words from each speech transcript. The resulting data was then lemmatized and tokenized.

\begin{table}[hbt!]
\caption{President Reagan's speech dataset}
\begin{center}
\begin{tabular}{ |c|c|c|c| } 
\hline
\textbf{Year} & \textbf{No. of speeches} \\ 
\hline
1980 & 6 \\
\hline
1981 & 8 \\
\hline
1982 & 11 \\
\hline
1983 & 12 \\
\hline
1984 & 14 \\
\hline
1985 & 13 \\
\hline
1986 & 14 \\
\hline
1987 & 10 \\
\hline
1988 & 9 \\
\hline
1989 & 1 \\
\hline
\end{tabular}
\label{tab:speeches}
\end{center}
\end{table}

\textbf{Part-of-speech (POS) features}: People diagnosed with AD use more pronouns than nouns. Therefore, POS features such as the number of pronouns and the pronouns-to-nouns ratio are important. We identified adverbs, nouns, verbs, and pronouns for each speech transcript with natural language processing (NLP) tools. POS tags having at least 10 occurrences were selected to compute their percentage ratios. Similarly, words that have at least a frequency of 10 were selected and their occurrence percentages were computed.  
The full set of \textit{POS features} we used were: (1) number of pronouns, (2) pronoun-noun ratio, (3) number of adverbs, (4) number of nouns, (5) number of verbs, (6) pro-noun frequency rate, (7) noun frequency rate, (8) verb frequency rate, (9) adverb frequency rate, (10) word frequency rate, and (11) word frequency rate without excluding stop words. 

\textbf{Vocabulary Richness}: AD patients show a decline in their vocabulary range. Therefore, vocabulary richness metrics: Honore’s Statistic (HS), Sichel Measure (SICH), and Brunet’s Measure (BM) were calculated for each speech. Higher values of Honore’s and Sichel measures indicate greater vocabulary richness. But a higher value corresponds to low vocabulary richness for the Brunet’s measure. 

\textbf{Readability Measures}: We computed two readability measures, namely, Automated Readability Index (ARI) and Flesch-Kincaid readability (FKR) score. A higher ARI indicates complex speech with rich vocabulary whereas lower Flesch-Kincaid score indicates rich vocabulary. 

\begin{figure}[hbt!]
\centering
\includegraphics[width=8cm,height=6cm]{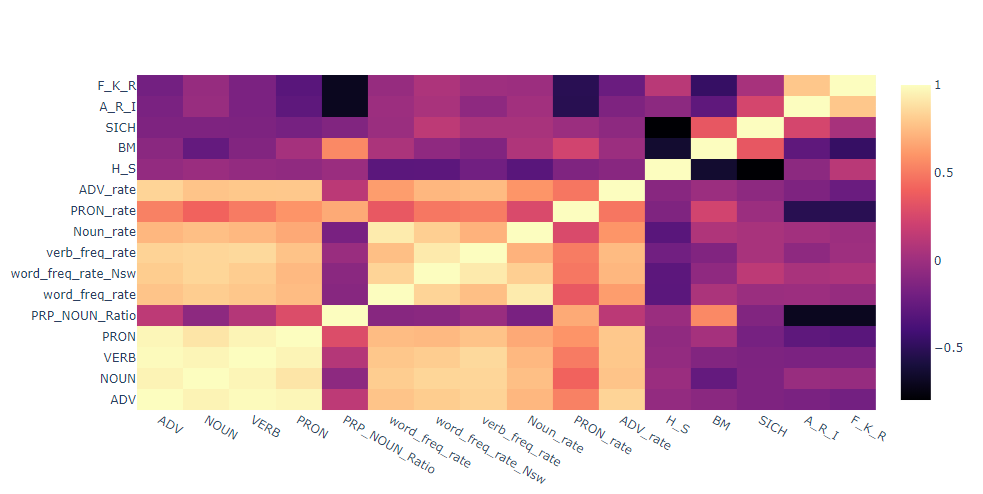}
\caption{Correlation matrix of the linguistic features.}
\label{fig:corrmat}
\end{figure}
Figure~\ref{fig:corrmat}
shows the correlation matrix of the chosen linguistic features computed for the Reagan speech dataset. Note that some of the features are highly correlated, therefore, we pruned the feature set to the following 9 features:
\begin{enumerate}
    \item pronoun-noun ratio
    \item word frequency rate 
    \item verb frequency rate
    \item pronoun frequency rate
    \item adverb frequency rate
    \item Honore’s measure
    \item Brunet’s measure
    \item Sichel measure
    \item Automated Readability Index
\end{enumerate}
Any further analysis in this paper uses the above 9 selected linguistic features.

\section{Clustering and Visualization of Linguistic Features}
\label{sect:clusterAndVisual}

We selected the t-SNE machine learning technique for clustering and visulaization of linguistic features extracted from Reagan’s speeches. t-SNE is better at creating a single map for revealing structures at many different scales important for high-dimensional data that lie on several different, but related, low-dimensional manifolds. This implies that it can capture much of the local structure of the high-dimensional data very well, while also revealing global structure such as the presence of clusters at several scales (van der Maaten and Hinton 2008). If there are AD-related linguistic patterns then t-SNE may reveal them as clusters. 

In t-SNE, the high-dimensional Euclidean distances between datapoints are converted into conditional probabilities representing similarities between them. For example, the similarity of data point $x_j$ with $x_i$ is the conditional probability $p_{j|i}$ that $x_i$ would pick $x_j$ as its neighbor. Neighbors are picked in proportion to their probability density under Student $t$-distribution centered at $x_i$ in the low-dimensional space. Then the Kullback-Leibler divergence between a joint probability distribution, P, in the high-dimensional space and a joint probability distribution, Q, in the low-dimensional space is then minimized:
\begin{equation}
\mathbf{\min}{KL}({P}||{Q}) = {\sum_i}{\sum_j}{p}_{i_j}{\log\frac{{{p}_{i_j}}}{{q}_{i_j}}}
\end{equation}

t-SNE has two tunable hyperparameters: perplexity and learning rate. Perplexity allows us to balance the weighting between local and global relationships of the data. It gives a sense of the number of close neighbors for each point. A perplexity value between 5 and 50 is recommended. The learning rate for t-SNE is usually in the interval [10, 1000]. For a high learning rate, each point will approximately be equidistant from its nearest neighbours. For a low rate, there will be few outliers and therefore the points may look compressed into a dense cloud. Since t-SNE’s cost function is not convex different initializations can produce different results.
After tuning t-SNE’s hyperparameters for the dataset, we chose perplexity value equal to 4 and learning rate equal to 100. 

\begin{figure}[hbt!]
\centering
\includegraphics[width = 70 mm, height = 50 mm]{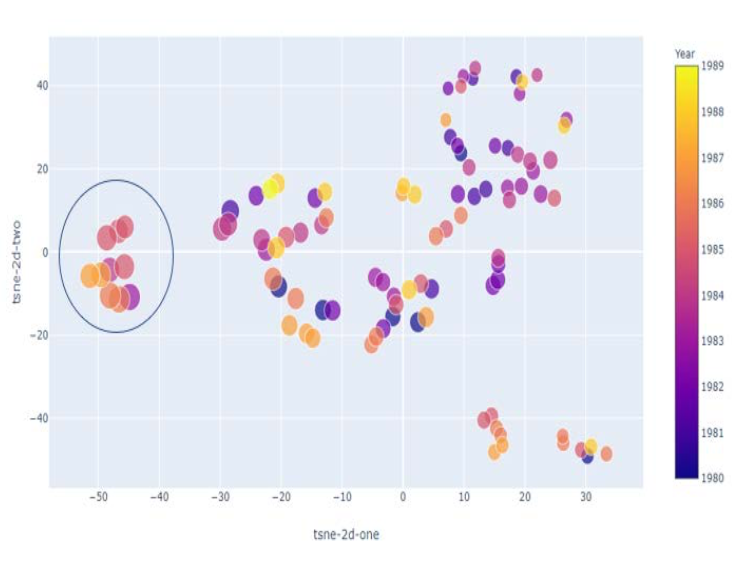}
\caption{ 2-dimensional visualization of speech transcripts where size of each circle in the map is proportional to the pronoun-to-noun ratio.}
\label{fig:fig4}
\end{figure}

\textbf{Pronoun-to-noun ratio}: Figure~\ref{fig:fig4} shows t-SNE based clustering of the speech transcripts. Here, the radius of each circle is proportional to the pronoun-to-noun ratio. Notice that the cluster on the left side of the graph contains speeches (from 1983 to 1987) have higher values of the ratio. Recall that higher pronoun-to-noun ratio is an indicator of early stage AD.

\begin{figure}[hbt!]
\centering
\includegraphics[width = 70 mm, height = 50 mm]{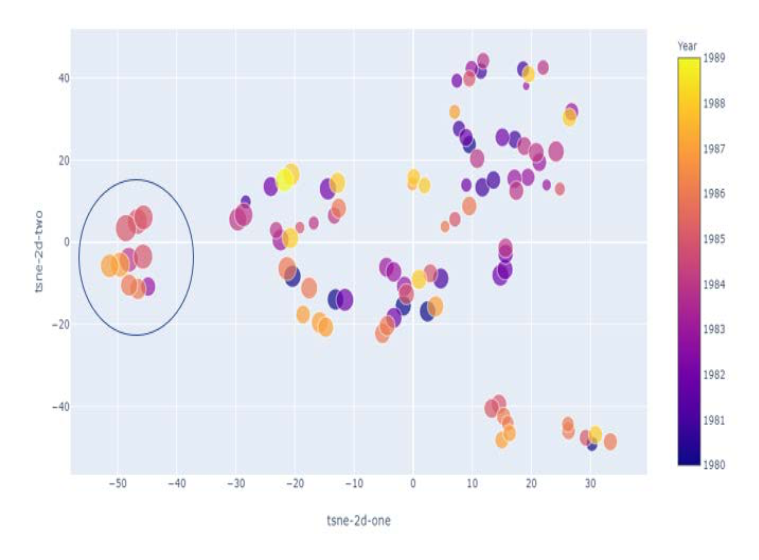}
\caption{ 2-dimensional visualization of speech transcripts where size of each circle in the map is proportional to the pronoun frequency.}
\label{fig:fig5}
\end{figure}

\textbf{Pronoun frequency}: Fig.~\ref{fig:fig5} indicates clustering and low-dimensional visualization results when the size of each circle is proportional to the pronoun frequency. We again see that the cluster on the left side of the map contains speeches with higher pronoun frequency, another indicator of early stage AD.

\begin{figure}[hbt!]
\centering
\includegraphics[width = 70 mm, height = 50 mm]{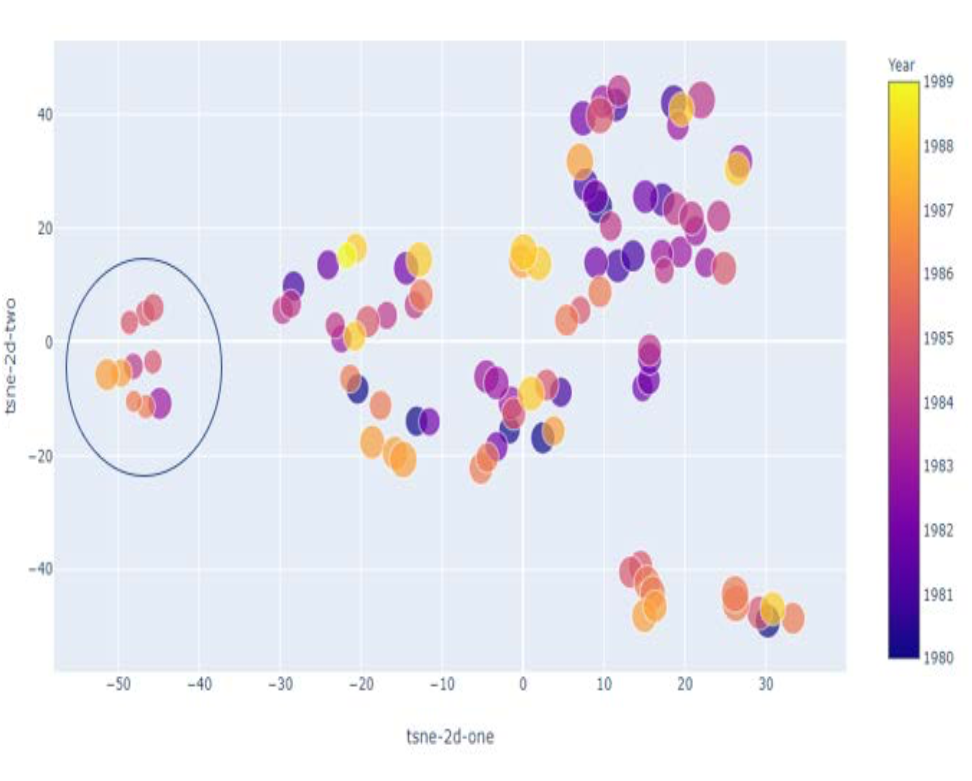}
\caption{ 2-dimensional visualization of speech transcripts where size of each circle in the map is proportional to the readability score.}
\label{fig:fig6r}
\end{figure}

\textbf{Readability score}: From Fig.~\ref{fig:fig6r} we observe that the speeches from 1983-1987 have lower readability scores.

\begin{figure}[hbt!]
\centering
\includegraphics[width = 70 mm, height = 50 mm]{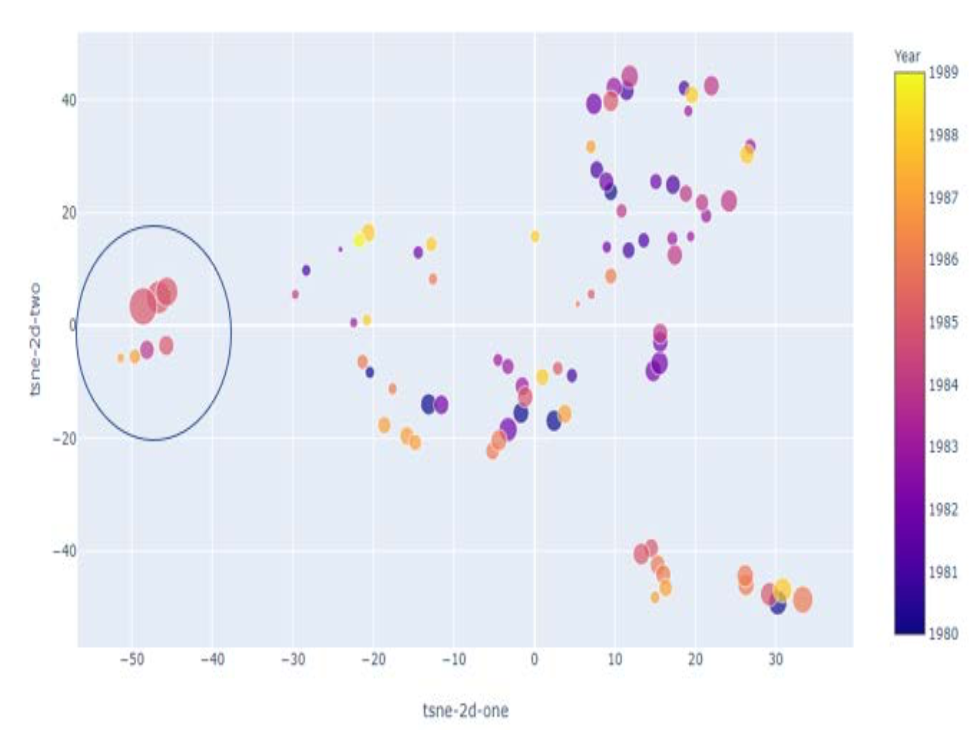}
\caption{ 2-dimensional visualization of speech transcripts where size of each circle in the map is proportional to the word repetition frequency.}
\label{fig:fig6}
\end{figure}

\textbf{Word repetition frequency}: The word repetition frequency map in Fig.~\ref{fig:fig6} shows that three speeches standout. These three speeches have a higher repetition of high frequency words. Interestingly, two of these speeches have word lengths smaller than the mean length of all the speeches.

From these clustering results the speeches identified as showing early signs of AD are: 
\begin{itemize}
    \item 03-23-1983: Address to the Nation on National Security (“Star Wars” SDI Speech)
    \item 05-28-1984: Remarks honoring the Vietnam War’s Unknown
    \item 02-06-1985: State of the Union Address, (American Revolution II)
    \item 04-24-1985: Address to the Nation on the Federal Budget and Deficit Reduction
    \item 05-05-1985: Speech at Bergen-Belsen Concentration Camp Memorial
    \item 05-05-1985: Speech at Bitburg Air Base
    \item 04-14-1986: Address to the Nation on the Air Strike Against Libya
    \item 06-24-1986: Address to the Nation on Aid to the Contras
    \item 08-12-1987: Address to the Nation on the Iran-ContraAffair
    \item 08-12-1987: Address to the Nation on the Iran-ContraAffair
    \item 09-21-1987: Address to the General Assembly of the
United Nations, (INF Agreement and Iran)
\end{itemize}

Therefore, t-SNE based clustering and low-dimensional visualization of President Reagan’s speeches from 1964 to 1989 reveals the following:
\begin{itemize}
    \item he started showing signs of early AD well before the official announcement in 1994
    \item it is highly likely that he developed AD sometime between 1983 and 1987
    \item over time, President Reagan’s showed a significant reduction the number of unique words but a significant increase in conversational fillers and non-specific nouns
    \item the proposed method identifies specific speeches that exhibit linguistic markers for AD
\end{itemize}
Some of these findings are corroborated by prior research that analyzed his interviews and compared them with President Bursh's public speeches, \cite{BerEtal15}. 

\section{Linguistic Anomaly Detector for AD}
\label{sect:adad}

t-SNE-clustering-based approach provides visualization of speeches that are statistically different (``anomalies"). But we need an automated method to identify these anomalies for early signs of AD. Therefore, we investigate a one-class support vector machine (SVM) anomaly detector \cite{erfani2016high}. This method is useful in practice when majority of a subject's speeches over several years would be (statistically) typical of a healthy control (``normal") until he/she begins to exhibit early signs of AD. Our hypothesis is that early stage AD will begin to reveal itself as  statistical anomalies in linguistic feature space. In this section we investigate this hypothesis.


We designed a one class SVM with the following hyperparameter values (the choices of these values are not discussed for the sake of clarity and focus), $\nu=0.5$ (an upper bound on the fraction of training errors and a lower bound of the fraction of support vectors), kernel=rbf (radial basis function) and $\gamma=\frac{1}{\mathrm{(number~of~features \times variance~of~features)}}$ (kernel coefficient). 
\begin{figure}[hbt!]
\centering
\includegraphics[width = 75 mm, height = 60 mm]{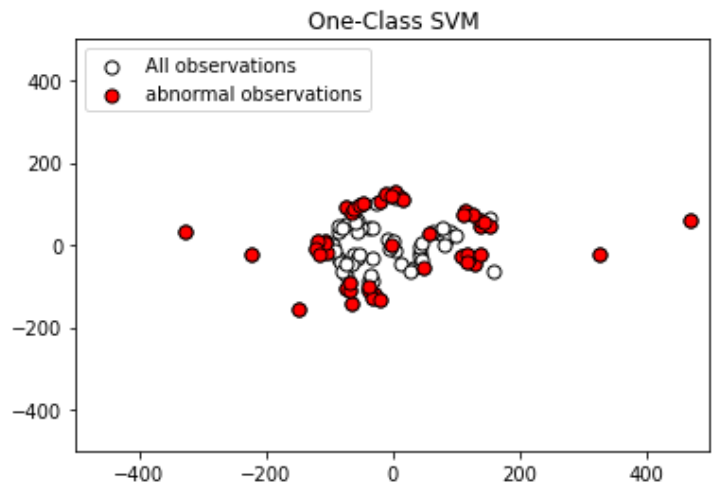}
\caption{Speeches from 1984 to 1986 are detected as abnormal or anomalies.}
\label{fig:fig7}
\end{figure}
Figure~\ref{fig:fig7} shows that the one class SVM detector identified several speeches from 1984 to 1986 as anomalous. That is, President Reagan's speeches in these years are different from the previous years in the linguistic feature space. Therefore, it is likely that: 
\begin{itemize}
    \item he started showing signs of early AD well before the official announcement in 1994
    \item the onset of AD start from 1984 and became more pronounced in 1985 and 1986, which is corroborated by prior research (e.g., \cite{BerEtal15}) that analyzed his interviews. 
\end{itemize}
 
One class SVM learns the profile of non-AD speeches as ``normal" over a period of time and detects anomalies to signal the onset of AD. But in many practical instances long historical data may not be available for a subject.
In this case, we must identify anomalies explicitly instead of learning what is normal. Isolation forest \cite{ding2013anomaly} is an unsupervised machine learning algorithm that is applicable for this purpose. 
\begin{figure}[hbt!]
\centering
\includegraphics[width = 75 mm, height = 60 mm]{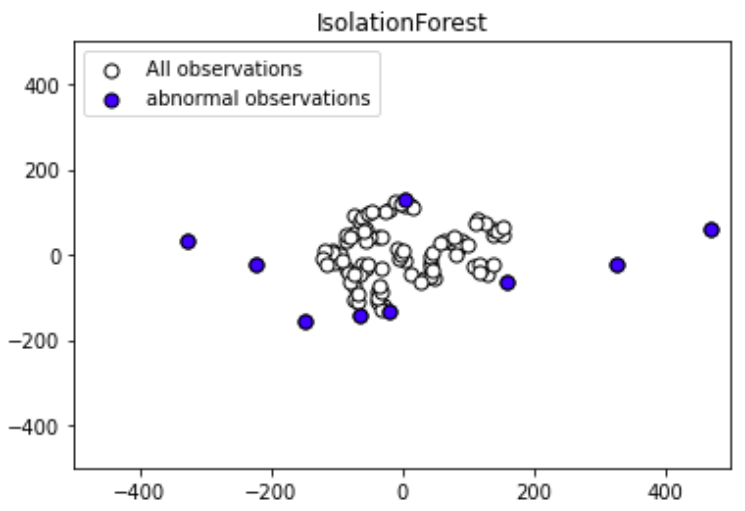}
\caption{ Isolation forest algorithm for AD detection.}
\label{fig:fig9}
\end{figure}

\begin{figure}[hbt!]
\centering
\includegraphics[width = 60 mm, height = 40 mm]{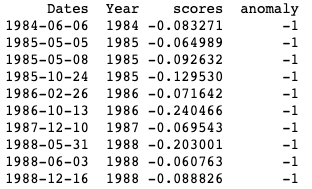}
\caption{Year-wise speeches identified as anomalous.}
\label{fig:fig10}
\end{figure}

We applied the isolation forest algorithm on the speech dataset and tuned its hyperparameters. Figure~\ref{fig:fig9} and Figure~\ref{fig:fig10} show the results. Figure~\ref{fig:fig9} shows the ten speeches that were isolated as anomalies. The corresponding dates of these speeches are seen in Figure~\ref{fig:fig10}. We observe that a few speeches between 1984 and 1988 as linguistic anomalies. This time period overlaps significantly with the results of once class SVM and t-SNE clustering.

\section{Conclusions}
\label{sect:conc}
 A set of nine linguistic biomarkers for AD was identified. Two complementary unsupervised machine learning methods were applied on the linguistic features extracted from President Reagan's 98 speeches given between 1980 to 1989. The first method, t-SNE, identified and visualized speeches indicating early onset of AD. A higher pronoun usage frequency, lower readability scores, higher repetition of high frequency words were revealed to be the key characteristics of potential AD-related speeches. A subset speeches from 1983 to 1987 were detected to possess these characteristics. 
 
 The second machine learning method, one class SVM, learned what is ``normal" (i.e., non-AD speech) to detect anomalies in speeches over a period of time. This approach detected several speeches between 1984 and 1986 as potential AD-related. Since normal speech may not be available historically we applied the isolation forest algorithm that explicitly detects anomalies without learning what is normal. This detected 10 speeches from 1983 to 1987 as AD-related. 
 
 From the experimental analysis our conclusion is that President Reagan had signs of AD sometime between 1983 and 1988. This conclusion corroborates results from other studies in the literature. Note that that President Reagan had AD was publicly disclosed only in November 1994.

\bibliography{references}

\bibliographystyle{unsrt}  
\end{document}